\documentclass[9pt,twocolumn,twoside]{osajnl}
\pdfoutput=1
\usepackage{subcaption,graphicx,trimclip}
\usepackage[nameinlink]{cleveref}
\journal{ol}
\setboolean{shortarticle}{true}
\usepackage[utf8]{inputenc}
\usepackage[english]{babel}
\usepackage[page]{appendix}
\usepackage{hyperref}
\usepackage{etoolbox} 
\usepackage{bm}
\usepackage{hyperref}
\usepackage{bold-extra}
\usepackage{tabularx}
\usepackage{float}
\usepackage{MnSymbol}
\usepackage{booktabs}
\usepackage{nicefrac} 
\usepackage[version=4]{mhchem}

\Crefformat{appendix}{Supplementary Information #2#1#3}

\newcommand{\mytitle}{Terahertz Voltage-controlled Oscillator from a Kerr-Induced Synchronized Soliton Microcomb}

\author[1,2,\dag]{Usman A. Javid}
\author[1,3,\dag]{Michal Chojnacky}%
\author[1,2, *]{Kartik Srinivasan}%
\author[1,2,*]{Gr\'egory Moille}%
\affil[1]{Joint Quantum Institute, NIST/University of Maryland, College Park, USA}
\affil[2]{Microsystems and Nanotechnology Division, National Institute of Standards and Technology, Gaithersburg, USA}
\affil[3]{Sensor Science Division, National Institute of Standards and Technology, Gaithersburg, USA}
\affil[$\dag$]{These authors contributed equally}
\affil[*]{Corresponding authors:kartik.srinivasan@nist.gov; gregory.moille@nist.gov}
\date{\today}

\begin{document}

\thispagestyle{fancy}

\date{\today}

\title{\mytitle}
\begin{abstract}
  The generation of controlled and arbitrarily tunable terahertz radiation, essential for many applications, has proven challenging due to the complexity of experimental setups and fabrication techniques. We introduce a new strategy involving control over a terahertz repetition rate integrated frequency comb, using Kerr-induced synchronization, that results in a terahertz-voltage-controlled oscillator. By modulating the reference laser, we can transfer any microwave waveform onto the microcomb repetition rate via a linear transfer function based on optical frequency division. The resulting frequency comb with a terahertz carrier can be created using integrated components, with a bandwidth constrained only by the synchronization bandwidth and high coherence resulting from the low-noise soliton microcomb in the Kerr-induced synchronized state.
\end{abstract}
\maketitle

\section{Introduction}
\begin{figure*}[t]
  \centering
    \includegraphics{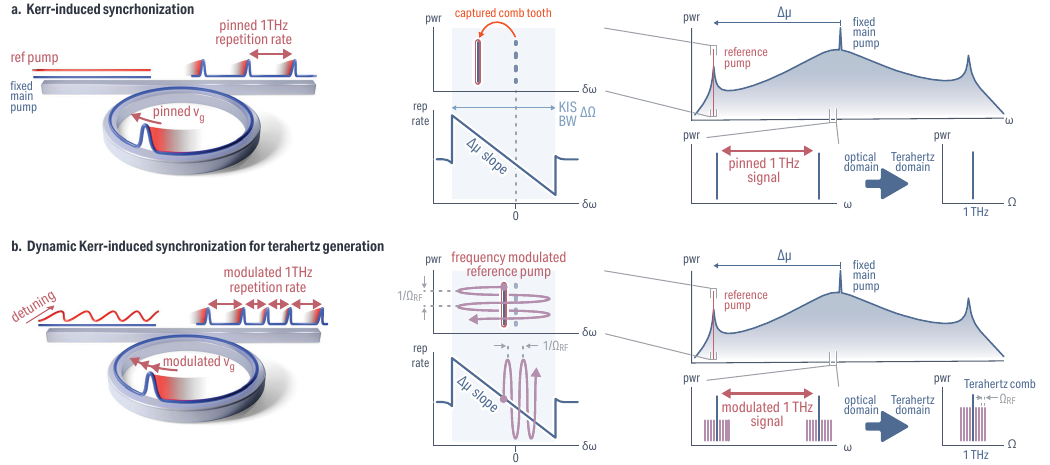}
    \caption{\label{fig:1}%
    \textbf{Kerr induced synchronization for dynamic comb repetition rate tuning for terahertz generation -- }%
    \textbf{a} Kerr-induced synchronization (KIS) starts with a dissipative Kerr soliton (DKS) traveling in a microring resonator, which is generated by a main pump laser. A reference laser is sent to the same microring resonator, whose circumference is set to produce a free spectral range near 1~THz. When this reference laser is sufficiently close to one comb tooth (within the KIS bandwidth), it will capture this comb tooth. The comb, already pinned by the main pump as a comb tooth, then becomes double-pinned. Since the KIS bandwidth is non-null, the reference can be tuned to change the comb repetition rate or can be fixed to lock the comb repetition rate. The repetition rate of the comb can be detected by sending two adjacent comb lines to a photomixer, creating a low-noise terahertz signal. 
    \textbf{b} Thanks to the large KIS bandwidth, sending a frequency-modulated reference laser into the microring resonator enables the repetition rate to be modulated at the same rate. As such, the terahertz signal detected by the photomixer from the beating of the now modulated two adjacent comb lines will present spectral components around the nominal repetition rate frequency. This process creates a coherent, low-noise terahertz comb that harnesses the intrinsic optical frequency division property of the frequency comb.
    }
\end{figure*}

Technological developments in the terahertz spectral region (0.1 THz to 10~THz) have recently expanded rapidly~\cite{DangNat.Electron.2020, ElayanIEEEOpenJ.Commun.Soc.2020, LeitenstorferJ.Phys.Appl.Phys.2023, MittlemanJ.Appl.Phys.2017}, bridging the so-called
``terahertz gap''. This range is of particular interest for a wide variety of applications, including molecular spectroscopy~\cite{BeardJ.Phys.Chem.B2002, JepsenLaserPhotonicsRev.2011}, remote sensing~\cite{LiuNaturePhoton2010}, medical imaging~\cite{YangTrendsinBiotechnology2016}, and climate monitoring~\cite{BornPlantPhysiology2014, WedageIEEENetw.2023}, while enabling the new generation of 6G wireless networks~\cite{NagatsumaNat.Photonics2016, RappaportIEEEAccess2019} to meet the ever-increasing demand for ultra-fast wireless communication.
Terahertz radiation may be generated in two distinct manners: electronically, by the up-conversion of microwave frequencies through electro-optic multiplication~\cite{SenguptaNat.Electron.2018}, or optically, by mixing two separated optical tones detected by an ultra-fast photomixer~\cite{ItoJ.LightwaveTechnol.JLT2005} or through the use of nonlinear crystals via optical rectification~\cite{HoffmannOpt.ExpressOE2007}. The noise associated with the radio frequency and optical waves directly impacts the quality of the generated terahertz signal, which can present challenges when low-noise operation is required. In such applications, an alternative approach involves harnessing the intrinsic low-noise properties of an optical frequency comb, which creates a phase-coherent link between the optical and microwave domains~\cite{DiddamsScience2020a}. A microwave signal can be extracted from photodetection of the comb repetition rate (or a harmonic of it), with low-noise performance realized through optical frequency division (OFD), where the repetition rate frequency noise is given by the mutual noise of two optically locked comb teeth divided by the integer comb tooth spacing ($\Delta \mu$) between them. OFD has been established as an effective method for ultra-low noise synthesis of microwave signals~\cite{DialloOpt.Lett.OL2017, FortierLaserPhotonicsRev.2016, FortierNaturePhoton2011, HamidiIEEETrans.Microw.TheoryTech.2010}. To extend OFD to the terahertz regime, scaling down the comb size is critical, as the comb repetition rate is inversely proportional to the cavity size. Such downscaling is feasible with a chip-integrated microring resonator made of $\chi^{(3)}$ nonlinear material, which supports dissipative Kerr soliton (DKS) generation~\cite{KippenbergScience2018}. The traveling DKS in the microring is periodically extracted at every round-trip, creating a pulse train output corresponding to a low-noise microcomb. When the microcomb's repetition rate is detected, it enables successful transduction from the optical domain to the millimeter wave~\cite{GreenbergAPLPhotonics2024, TetsumotoNat.Photonics2021, TetsumotoOpt.Lett.2020}. As such soliton microcombs can also be battery operated~\cite{SternNature2018b}, the size, weight, and power (SWaP) of this terahertz generation approach can be compelling. 

While terahertz synthesis via microcombs presents a promising pathway, on-chip control and tuning of the terahertz repetition rate of the DKS remain challenging. Approaches that rely on changing the properties of the microring resonator, through thermo-refractive~\cite{MoilleAPLPhotonics2022a, NiuNatCommun2023}, electro-optic~\cite{HeNat.Commun.2023}, or piezoelectric~\cite{LiuNature2020} elements, enable limited repetition rate tuning while mostly impacting the cavity resonance frequencies, with the latter forcing the main pump to be re-tuned for maintaining the soliton state. 
In contrast, a new approach has recently been introduced, termed Kerr-induced synchronization (KIS)~\cite{MoilleNature2023}, which enables all-optical tuning of the DKS repetition rate. Here, a secondary reference laser is injected into the microring and synchronizes the cavity soliton. Under the correct conditions, such synchronization results in the capture by the reference laser of the nearest comb tooth, with tuning of the reference laser within the KIS bandwidth resulting in direct control of the DKS repetition rate. Thanks to the significant KIS bandwidth (up to a few gigahertz), this enables DKS repetition rate modulation. This approach solely modifies the DKS, without affecting the resonator, making high-speed modulation while maintaining the low-noise DKS state achievable.
In this work, we demonstrate that KIS can be used to create a terahertz voltage-controlled oscillator through voltage-controlled modulation of the reference laser frequency. We show that the modulation of the reference laser frequency is directly transduced onto the DKS repetition rate via OFD, enabling the generation of low noise arbitrary waveforms centered at around 1~THz. Our results suggest that terahertz generation via KIS of DKS microcombs can be used in applications such as spectroscopy, ranging, and imaging.

\begin{figure*}[t]
  \centering
    \includegraphics{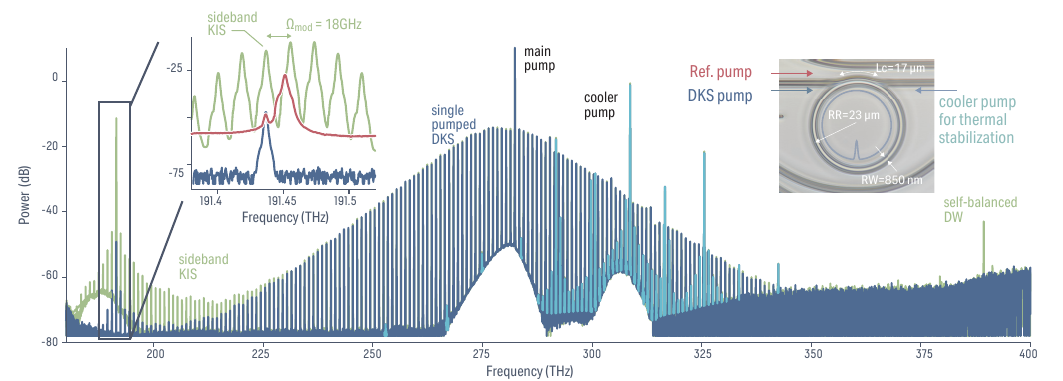}
    \caption{\label{fig:2}%
    \textbf{1~THz octave spanning frequency with sideband Kerr-induced synchronization -- }%
    The microring used (inset left) is made of \ce{Si3N4} with a ring width of 860 nm and an outer ring radius of 23 {\textmu}m embedded in silica. The counter-propagating and cross-polarized cooler pump, relative to the main pump at 283~THz, helps thermally stabilize the resonator for adiabatic access to the DKS state, while minimizing its nonlinear interaction with the soliton (teal spectra). The comb exhibits a DW at 191.5~THz (blue spectrum) which we leverage for energy-efficient Kerr-induced synchronization (green spectra). Contrary to previous work, we modulate the reference laser at $\Omega_\mathrm{mod} = 18$ GHz and offset it from the comb tooth by the same amount for the lower frequency sideband to achieve synchronization (see left inset). This approach allows the reference to remain fixed and potentially locked to a stable cavity for low-noise operation, while enabling sideband frequency modulation with minimal residual amplitude modulation. Once the sideband is synchronized, the same repetition rate modulation scheme can be applied. A similar phenomenon can be observed for the short DW at 389~THz, which is amplified through self-balancing~\cite{MoilleFront.Opt.2023}.
    }
  \end{figure*}
\section{Concept}
  First, it is important to recall the fundamental properties of KIS and understand how it can be harnessed to create tunable terahertz oscillators [\cref{fig:1}a]. A continuous wave (CW) main pump creates a DKS inside a ring resonator cavity. By also injecting a weak reference pump laser, the $\chi^{(3)}$ nonlinearity of the resonator enables the DKS and the intracavity reference laser to bind their group velocities through cross-phase modulation (XPM)~\cite{WangOptica2017a, Moille2024_colorKIS}. When the reference laser frequency is close enough to a comb tooth that is $\mu_s$ modes away from the main pump, the XPM is sufficient for the soliton and reference to also synchronize their phase velocities~\cite{MoilleNature2023}. This synchronization corresponds to the closest comb tooth snapping onto the reference laser, making them indistinguishable. %
  Since the comb is now doubly pinned by the main pump and the reference pump -- fixing the group velocity of the traveling DKS in the resonator -- the repetition rate noise of the comb is converted from the pump lasers' noise via OFD~\cite{MoilleNature2023,WildiAPLPhotonics2023}. These two pumps can be optically locked to cavity references, enabling an ultra-low noise repetition rate. Since the DKS round trip time can be made to be in the picosecond regime, a high quality terahertz signal can be generated by detecting the beating of two adjacent comb lines. %
  One important property of KIS is its bandwidth $\Delta \Omega$, determined by the KIS coupling energy, which is the geometric mean of the intracavity energy of the reference and synchronized comb tooth, respectively~\cite{MoilleNature2023}. Therefore, the synchronization happens within a frequency window around the comb tooth, for which the reference will capture this tooth. Hence, the soliton must adapt its group velocity -- thus the comb repetition rate -- for the comb tooth captured by the reference to still be on a fixed frequency grid. Along with the potential for better phase matching to improve dispersive wave (DW) generation~\cite{MoilleFront.Opt.2023}, KIS enables direct frequency modulation of the reference pump to be coherently transferred onto the comb repetition rate~[\cref{fig:1}b]. In this regime, the DKS acts as a transducer where the up-shifted microwave signal at $\Omega_\mathrm{RF}$ is imprinted on the reference frequency modulation, via a voltage controlled oscillator (VCO) combined with an electro-optic modulator, which is then frequency down-shifted to the terahertz domain.
  The system therefore becomes analogous to a frequency modulator, where a frequency comb in the terahertz domain can be generated, where each comb line $n$ around the repetition rate carrier frequency will be separated by $\Omega_\mathrm{RF}$ and will exhibit an amplitude $A_n= J_n(\frac{\delta \Omega}{\Omega_\mathrm{RF}})$, with $J_n$ being the $n$\textsuperscript{th} Bessel function. The modulation amplitude $\delta \Omega$ of the repetition rate $\omega_\mathrm{rep}$, which is directly linked to the modulation amplitude of the reference by a factor $\mu_s$, leads to the terahertz comb and can have a maximum value of half of the KIS bandwidth $\Delta \Omega$ divided by the OFD factor $\Delta\mu$, and will define the maximum possible excursion of the modulation, while the inverse of the cavity photon lifetime will determine the maximum speed at which the modulation $\Omega_\mathrm{RF}$ can be applied.


\section{Results}
\subsection{Optical characterization of the microcomb}
To experimentally study this effect, we use a $H=670$~nm thick silicon nitride (\ce{Si3N4}) microring resonator with a ring width $RW=860$~nm and an outer ring radius of $RR=23$~{\textmu}m, where the DKS round-trip time is around 1~ps, hence creating a $\omega_\mathrm{rep}/2\pi \approx 1$~THz frequency comb repetition rate. The dispersion of the fundamental transverse electric mode of this ring is engineered such that for a pump in the 283~THz range (1061~nm range), the dispersion is anomalous with additional higher order dispersion terms allowing for phase matching around 191.5~THz (1566~nm) and 389~THz (770~nm), for which the DKS can emit DWs [\cref{fig:2}], creating an octave spanning comb~\cite{MoilleNature2023, YuPhys.Rev.Applied2019a}. To generate the soliton, we use a counterpropagating cooler pump in the transverse-magnetic polarization to minimize the nonlinear coupling with the main pump while enabling temperature stabilization of the resonator for adiabatic DKS access with 140~mW of on-chip main pump power~\cite{ZhangOptica2019a, ZhouLightSciAppl2019}. The short DW is not properly extracted in the output bus waveguide since the coupling dispersion using a $L_c = 17$~{\textmu}m pulley-like coupler~\cite{moille_broadband_2019} has been engineered mostly for balancing the critical coupling regime at frequencies between the main pump down to the lower frequency DW at 191.5~THz.
To trigger KIS, we send a weak reference laser pump close to the maximum of the low frequency DW. The advantage of this choice is two fold: the comb tooth being per definition close to a cavity mode resonance, the reference intracavity field can be maximized, while the resonance also enhances the comb tooth energy. Consequently, the KIS coupling energy $E_\mathrm{kis} = \sqrt{\frac{\kappa_\mathrm{ext}}{\kappa^2}P_\mathrm{ref}E_{\mu s}}/E_\mathrm{dks}$ is maximized~\cite{MoilleNature2023,Moille2024_ACKIS}, with $\kappa_\mathrm{ext}=2\pi \times 100$~MHz, $\kappa = 2\kappa_\mathrm{ext}$ the total loss rate ($\kappa_\mathrm{ext}$ is the waveguide coupling rate), $P_\mathrm{ref}$ the on-chip reference pump power, $E_{\mu s}\approx  1.2 \times 10^{-4}$~J the energy of the comb tooth at the synchronized mode (whose index $\mu_s=-91$ is with respect to the main pump), and $E_\mathrm{dks}\approx 10^{-2}$~J the intracavity energy of the DKS obtained from the Lugiato-Lefever equation~\cite{MoilleJ.RES.NATL.INST.STAN.2019}. The KIS bandwidth being directly related to the KIS energy such that $\Delta \Omega = 4\mu_s\omega_\mathrm{rep}E_\mathrm{kis}$, working at the DW enables us to obtain a large KIS bandwidth of 1.2~GHz with $P_\mathrm{ref} \approx 200$~{\textmu}W sufficient to obtain a large enough frequency modulation contrast. 
Although KIS using direct modulation of the laser frequency (through current modulation) is possible, the targeted excursion range of a few gigahertz will also induce amplitude modulation. Though KIS should be robust against resulting variations in the reference energy, residual amplitude modulation will pose a challenge when locking to a reference cavity. Instead, here we propose to use an electro-optic modulator to create modulation sidebands at a large frequency separation $\Omega_\mathrm{mod} = \pm18$~GHz from the reference laser. By positioning the reference laser at $\approx \Omega_\mathrm{mod}$, one of the modulation sidebands can synchronize the soliton~[\cref{fig:2}]. Such a large frequency separation between the reference and its sideband is necessary to prevent other types of synchronization dynamics that arise if $\Omega_\mathrm{mod}$ is comparable to $\Delta\Omega$~\cite{Moille2024_ACKIS}; instead, setting $\Omega_\mathrm{mod} \gg\Delta\Omega$ allows us to treat the reference and its sideband as independent oscillators in the context of DKS synchronization. This approach allows for direct radiofrequency control of the sideband frequency that synchronizes DKS, while maintaining a fixed reference laser that can be locked to a cavity reference to reduce its linewidth and the resulting DKS repetition rate noise via OFD. For the remainder of the manuscript, we will refer to this reference sideband simply as the \textit{reference}, given that the reference carrier from which it is derived will remain unchanging and its contribution to the nonlinear dynamics of the system can be neglected. We adjust the reference laser power for the synchronization sideband to have an on-chip power of $200\ \mu W$.

\begin{figure}[!t]
  \centering
    \includegraphics{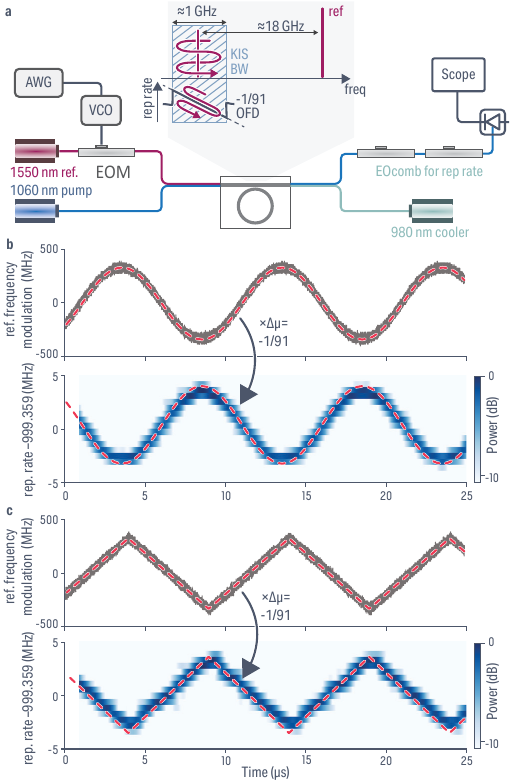}
    \caption{\label{fig:3}
    \textbf{Instantaneous repetition rate modulation characterization -- }%
    \textbf{a} Experimental setup schematic. The reference laser is modulated at $\Omega_\mathrm{mod}+\Omega_\mathrm{RF}$ using an arbitrary waveform generator (AWG) that drives a voltage control oscillator (VCO), with the output fed to an electro-optic phase modulator to create and control the sideband reference that synchronizes the comb. The repetition rate is detected using an electro-optic comb (EOcomb) apparatus, with the output beat note detected by an avalanche photodiode (APD) and measured with a fast oscilloscope. 
    \textbf{b} The sinusoidal AWG signal (top) at 100~kHz is transferred onto the reference through electro-optic modulation, and is in turn transduced onto the $\approx$999.35901 GHz repetition rate carrier through KIS (bottom). There are $\Delta\mu =-91$ comb teeth between the main pump and the reference, so the reference frequency excursion is divided onto the repetition rate by -91. The red lines represent the AWG average (top), which is optical frequency divided (OFD'd) onto the repetition rate (bottom), highlighting the linear transfer of the KIS transduction.
    \textbf{c} Similar to the sinusoidal AWG in (b), except with a triangular signal. The resolution of the sharp edge, which contains higher Fourier frequencies, highlights the large modulation bandwidth of the KIS transduction and its linearity.
    }
  \end{figure}

  \begin{figure*}
    \centering
      \includegraphics{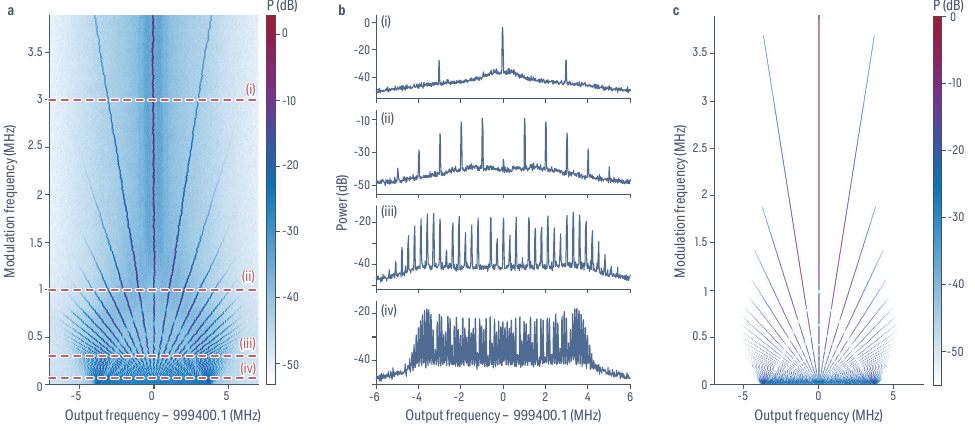}
      \caption{\textbf{Terahertz frequency-comb generation.} %
      \textbf{a} Spectrum of the modulated DKS repetition rate (measured with the EOcomb apparatus), hence the terahertz comb, through direct actuation via KIS with a modulated reference, for a modulation amplitude of $\delta\Omega=\mu_s^{-1}\times 341.4$ MHz, with $\mu_s =-91$ the mode at which KIS happens. The $y$ axis corresponds to the modulation frequency $\Omega_{\mathrm{RF}}$ ranging from 10~kHz to 4~MHz, hence impacting both the modulation excursion and the spacing between the terahertz comb teeth and their central carrier. The DKS repetition rate, hence the terahertz comb carrier, is at 999.40016 GHz \textpm 10~kHz, uncertainty from the $\omega_\mathrm{rep}$ linewidth  with free runing pumps~\cite{MoilleNature2023}, while the different spectra are recorded at a bandwidth of 100~Hz. %
      \textbf{b} Individual terahertz comb spectra for modulation frequencies $\Omega_\mathrm{RF} = 100$~kHz (iv), $300$~kHz (iii), $1$~MHz (ii), and $3$~MHz (i) as indicated by the dashed lines in a. The number and amplitude of terahertz comb teeth increases with $\Omega_\mathrm{RF}$ decreasing, since $\nicefrac{\delta\Omega}{\Omega_\mathrm{RF}}$ increases from (i)-(iv), with from $\nicefrac{\delta\Omega}{\Omega_\mathrm{RF}}= 114$ (i); $341$ (ii); $1138$ (iii); $3414$ (iv). %
     \textbf{c} Theoretically obtained terahertz comb spectrum at the same modulation frequencies and amplitudes as the experiment in a, using the experimentally calibrated response of the VCO (see supplementary material).}
      \label{fig:4}
    \end{figure*}
  
\subsection{Instantaneous response of the repetition rate}
We then study the modulation of the instantaneous repetition rate through the dynamic KIS. To modulate the reference (sideband), we source a voltage controlled oscillator (VCO) with an arbitrary waveform generator (AWG) to produce the microwave signal that drives an electro-optic phase modulator at $\Omega_\mathrm{mod}+\Omega_\mathrm{RF}$. This apparatus controls the sideband frequency and synchronizes the DKS comb tooth [\cref{fig:3}a]. We calibrate the transfer function of the AWG+VCO onto the modulation of the reference to account for the VCO's exhibited nonlinear response with frequency (see the supplementary material for more information). The output repetition rate is measured using two cascaded electro-optic phase modulators~\cite{DrakePhys.Rev.X2019a, MoilleNature2023, Moille2024_colorKIS, Moille2024_ACKIS, StonePhys.Rev.Lett.2020}, both modulated at 17.84706~GHz, enabling the overmodulation of two adjacent Kerr comb lines and forming an electro-optic comb (EOcomb), which can extract the DKS repetition rate. We use a narrow filter to select only the closest translated Kerr comb teeth, separated by less than 50~MHz, which can be detected by an avalanche photodiode (APD). From the beat we calculate $\omega_\mathrm{rep} = \pm\Omega_\mathrm{rep} + N\times \omega_{EO}\approx 2\pi\times 1$~THz, where $\Omega_\mathrm{rep}$ is the detected beat of the two translated Kerr comb teeth, $\omega_\mathrm{EO}$ is the EOcomb driving frequency, and $N=56$ is the number of of EOcomb teeth, which can be accurately measured with a $2\times 10^{-8}$ digit precision (\textit{i.e.,} 10~kHz) when the DKS is synchronized with the main and reference pumps free running, according to their respective laser linewidths and the OFD factor $\Delta\mu$ (locking the lasers enables $10^{-12}$ precision~\cite{MoilleNature2023}). Importantly, we measure the DKS repetition rate far from the reference portion of the comb spectrum around 279~THz (1075~nm) to avoid cross-talk with the reference modulation, while the main pump is kept fixed yet unlocked (\textit{i.e.,} free running). 

We temporally record the signal from the APD with a fast real-time oscilloscope with a 40~ps sampling interval, from which we reconstruct a spectrogram of the instantaneous repetition rate frequency against time for a 100~kHz sinusoidal microwave modulation [\cref{fig:3}b] and for a triangular modulation [\cref{fig:3}c] (see~\cite{Moille2024_ACKIS} for details on the experimental apparatus). We observe that the repetition rate follows the driving waveform accurately around its nominal carrier frequency at $\omega_\mathrm{rep}=999.35901$~GHz \textpm 10~kHz. After applying the expected OFD factor $\Delta\mu =-91$ corresponding to the comb tooth distance between the the main pump and the reference, it is clear that the repetition rate closely follows the microwave drive, demonstrating the faithful transduction into the millimeter-wave regime thanks to KIS of the DKS.  We note that the spectrogram resolution is only limited by the sampling rate/memory buffer size of the fast oscilloscope forcing a trade-off between temporal and spectral resolution. This can limit resolution of temporally fast and spectrally narrow responses, like the $C^1$ discontinuity corners when driven by the triangular AWG signal~[\cref{fig:3}c]. Yet, such a waveform is accurately reproduced even at the discontinuous change of slope  -- which contains higher Fourier frequences than the 100~kHz pattern frequency due to the nonlinearity of the response at that particular time -- highlighting the fast response time of the KIS transduction and its linearity. To understand the limits at which KIS can modulate the repetition rate, it is first important to recall that group velocity locking happens in low-dispersion systems through a simple XPM interaction without comb tooth capture, resulting in a mutli-color soliton system~\cite{WangOptica2017a, MoillearXiv2023}. On the other hand, KIS is a phase-locking mechanism in addition to the previously described group velocity locking (hence, the term \textit{synchronization}). The phase velocities of the DKS and the reference lock, making them indistinguishable~\cite{MoilleNature2023,Moille2024_ACKIS}. Their respective carrier envelope offset frequencies align while exhibiting the same repetition rate, and hence the reference captures its closest comb tooth~\cite{MoilleNature2023,WildiAPLPhotonics2023}.  This occurs since the phase slip time between the soliton and the intracavity reference $\tau_\mathrm{slip} = 2\pi c\left(\frac{1}{\omega_\mathrm{ref}} + \frac{1}{\omega_\mathrm{pmp}}\right)\frac{1}{\Delta v_\varphi} \approx 140$~ps is longer than the characteristic nonlinear time $\tau_\mathrm{NL} = \nicefrac{2\pi}{\gamma L |A|^2}\approx 30$~ps, with $\gamma = 2$~W\textsuperscript{-1}$\cdot$m\textsuperscript{-1} the effective nonlinearity in our system, $\Delta v_\varphi$ the mismatch of phase velocity, $\omega_\mathrm{pmp,ref}$ the frequency of the pump and reference, respectively, and $|A|^2$ the DKS peak power obtained from the Lugiato-Lefever equation~\cite{MoilleJ.RES.NATL.INST.STAN.2019}. Hence, the nonlinearity is fast enough to phase lock the reference and the DKS and yield KIS. It is important to note that the change of repetition rate for the KIS-modulated reference essentially occurs because the main pump is (loosely) pinned, and hence a change of carrier-envelope offset frequency $\omega_\mathrm{ceo}$ also changes $\omega_\mathrm{rep}$ according to the OFD factor $\Delta\mu$~\cite{Moille2024_ACKIS}. Once an intracavity variation of the reference occurs, the modification of the repetition rate will be distributed onto the DKS at a timescale $\tau_\mathrm{NL}$. However, the reference being an external field, the cavity is loaded on a timescale of the photon lifetime $\tau_\mathrm{phot} = \nicefrac{1}{\kappa}\approx 1$~ns $\gg \tau_\mathrm{NL}$. Hence, $\tau_\mathrm{phot}$ ultimately limits the speed of the modulation of the reference to actuate the comb repetition rate. Here, this limit is about 1~GHz, which is far faster than the triangular modulation frequency applied in~\cref{fig:3}b. Interestingly, to speed up the maximum modulation frequency, the cavity photon lifetime could be engineered such that the main pump remains in a critical coupling regime to optimize DKS generation while the reference mode could be overcoupled. Here the pulley coupling scheme allows for a quasi-dispersion-less coupling over the octave for which the comb spans~\cite{moille_broadband_2019}, balancing the coupling rate of both pumps. However, a much simpler straight bus coupling scheme would naturally allow for stronger coupling at the reference (longer wavelength) than at the pump. This would enable reduction of the quality factor by order of magnitudes (and correspondingly the photon lifetime at the reference), while also increasing the KIS energy coupling $E_\mathrm{kis}$ and the resulting synchronization bandwidth~\cite{Moille2024_ACKIS}.

\begin{figure}[t]
  \centering
    \includegraphics{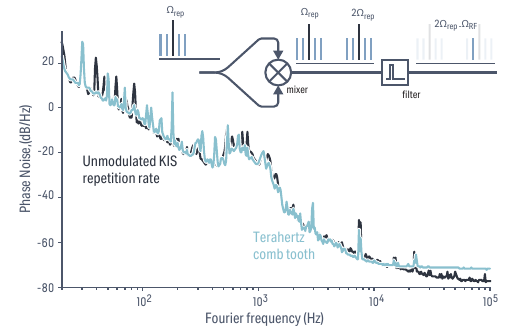}
    \caption{\label{fig:5}\textbf{Terahertz comb tooth noise measurement.} %
    Comparison of the DKS repetition rate phase noise (hence the terahertz comb carrier phase noise) in the static KIS regime (dark blue) and the noise of one of the terahertz comb teeth in the dynamic KIS regime of repetition rate modulation (teal). Their close overlap demonstrates the coherence of the produced terahertz comb through dynamic KIS repetition rate modulation. The terahertz comb tooth noise is measured using a homodyne technique, where the frequency down-shifted repetition rate $\Omega_\mathrm{rep}$ from the electro-optic comb apparatus, through filtering, allows us to measure the first negative terahertz comb tooth (inset). The AWG and VCO noise are negligible compared to the terahertz comb noise (about -120 dB/Hz at 1~kHz), since the experimental apparatus uses free-running lasers to synchronize the DKS. Further improvement of the terahertz comb noise can be readily available from optically locking the pump and reference laser to reference cavities. The power is referenced to that of the carrier, namely, dBc/Hz}
  \end{figure}

\subsection{Terahertz comb demonstration}
The sinusoidal frequency modulation of the repetition rate is interesting in the spectral domain as it leads to the formation of narrowly-spaced frequency combs with a carrier frequency at the Kerr optical frequency comb repetition rate $\omega_\mathrm{rep}$, thus leading to a terahertz comb. To investigate this, we characterize the obtained repetition rate-modulated frequency comb using a real-time electronic spectrum analyzer (RSA). We then scan the VCO frequency from $\Omega_\mathrm{RF}=$10~kHz to 4~MHz with a modulation amplitude $\delta\Omega=341.4/\Delta \mu$ ~MHz [\cref{fig:4}a], which is much smaller than the $\Delta\Omega = 1.2/\Delta \mu$~GHz KIS bandwidth once projected onto the repetition rate, allowing us to work in the same linear regime as previously described. At low frequencies, narrowly-spaced lines are observed, which space apart from the center as the modulation frequency increases. These lines are equally spaced at integer multiples of the modulation frequency from the carrier as expected from the spectrum of a frequency modulated sinusoid. At low frequencies, we can clearly see formation of a terahertz frequency comb with a center frequency of 999.40016 GHz \textpm 10~kHz [\cref{fig:4}b]. This terahertz comb exhibits reduced comb line power as $\Omega_\mathrm{RF}$ decreases, since reducing $\Omega_\mathrm{RF}$ not only reduces the repetition rate of the obtained comb, but also changes the modulation strength, thus impacting the terahertz comb span and comb line amplitudes. This is expected since, as previously described, the  terahertz comb tooth of order $N$ from the carrier frequency follows the well-known Bessel function relationship $A_n= J_n(\frac{\delta \Omega}{\Omega_\mathrm{RF}})$. Hence, for large modulation frequency, the spacing between terahertz comb teeth is obviously larger, but these comb teeth also present a lower amplitude [\cref{fig:4}b-i]. The terahertz comb becomes more dense with teeth whose amplitude increases with reduced $\Omega_\mathrm{RF}$~[\cref{fig:4}b ii to iv]. Interestingly, since the first order Bessel function $J_0$ presents zeros, one could also suppress the terahertz comb carrier, essentially removing the frequency component of the DKS repetition rate corresponding to the unmodulated KIS. For terahertz comb generation applications, the ability to arbitrarily control the bandwidth of the comb and its line spacing is a critical technical point compared to other techniques. Indeed, generation of terahertz frequency combs often requires complex processes such as mode-locking in quantum cascade lasers~\cite{BarbieriNaturePhoton2011}, or heterodyning a femtosecond laser at a terahertz photomixer \cite{YasuiApplPhysicsLett2006}. In DKS combs (without KIS), terahertz comb generation can be realized by post-processing the Kerr comb. For instance, we could pass one of the Kerr comb lines through a modulation circuit before beating it with an adjacent Kerr comb tooth or we could beat two Kerr combs with slightly different repetition rates, as done in dual-comb spectroscopy. In contrast, KIS allows modulation of the repetition rate within the resonator using an all-optical scheme with off-the-shelf components to create an arbitrary comb profile while leveraging the low-noise characteristics of DKS-based optical frequency combs. Furthermore, a terahertz comb obtained by beating two combs or lines of one comb spaced at multiples of the repetition rate \cite{YasuiApplPhysicsLett2006} will only yield a discrete set of accessible terahertz frequencies which could form a comb, but would not allow the continuous tuning expected from a terahertz VCO for arbitrary frequency control. To demonstrate the linear and continuous tuning of the repetition rate, we theoretically model the spectrum expected from a sinusoidally-modulated carrier signal (Fig.~\ref{fig:4}(c)). Here, we have plotted the power in each harmonic of the terahertz comb given by the previously described Bessel function model. By fitting the experimental data for $\delta\Omega_{\mathrm{rep}}$, and calibrating the VCO excursion $\delta\Omega$, we find that it agrees with the OFD factor $\Delta\mu=$-91. In this calculation we have also accounted for the limited temporal response of the driving electronics. The VCO has a response limited to a modulation frequency of 0.5~MHz, after which its excursion significantly drops. We have calibrated the frequency dependence of this excursion and applied a frequency dependent correction factor in the model in \cref{fig:4}c, with a detailed description and relevant data provided in the supplementary material. The results match the experimental data in Fig.~\ref{fig:4}a well, indicating that the DKS repetition rate has a purely sinusoidal modulation, demonstrating the linearity of the reference laser frequency transduction onto the repetition rate. 

\subsection{Terahertz comb noise characterization}
As described previously, the method we propose for generation of a terahertz comb is highly flexible compared to other methods, while also harnessing the low-noise repetition rate of the DKS. To support the latter point, we must also characterize the terahertz comb tooth noise. To do so, we homodyne the output of the APD with a mixer, creating up- and down-shifted copies of the spectrum (from $\Omega_{\mathrm{rep}}$ to DC and 2$\Omega_{\mathrm{rep}}$) [\cref{fig:5}]. The down-shifted spectrum is quite weak since the mixer typically requires a stronger local oscillator and the resulting signal-to-noise ratio is insufficient for detection by a phase noise analyzer (PNA). We filter out one comb line adjacent to the spectral center ($2\Omega_{\mathrm{rep}}-\Omega_{\mathrm{RF}}$) in the up-shifted spectrum using a software-based band-pass filter and amplify it. The phase noise of this comb line is measured by referencing against the AWG's internal clock signal at 10~MHz using a PNA. The generated noise curve is scaled down by 6~dB due to the frequency doubling that also doubles repetition rate noise present in all comb teeth (hence quadrupled in the electrical domain) [\cref{fig:5}]. We obtain a phase noise of -25 dB/Hz at 1~kHz and -68 dB/Hz at 10~kHz offset frequencies (the power is referenced to that of the carrier, namely, dBc/Hz). We also compare the terahertz comb tooth noise with the noise of the repetition rate in the KIS regime with the modulation turned off, referenced to the same clock signal. The plots show that the terahertz comb teeth carry the repetition rate noise as both noise curves overlap quite well. This demonstrates that the dynamic regime of KIS did not introduce any additional noise. Furthermore, since the DKS repetition rate is the dominant noise source, existing methods to lock and stabilize the DKS can be put to use to reduce the noise in the terahertz comb, for example see~\cite{LiuOpticaOPTICA2022a}.

                                              
\section{Discussion}
Since we have demonstrated highly flexible terahertz comb generation with coherence and potential for ultra-low noise operation once the two lasers that optically pin the OFC are locked, it is important to discuss and compare our approach with the current state-of-the-art, particularly those works leveraging integrated microcombs for tunable terahertz signal generation. Until now, DKS repetition rate control has mainly been exhibited through three effects: thermal tuning to modify the cavity refractive index via the thermo-refractive effect, piezo actuation for modulation of the cavity's physical dimensions, and electro-optic modulation using $\chi^{(2)}$ compatible materials. The first option through thermal tuning, in particular with integrated heaters, has been shown to tune the repetition rate significantly, from hundreds of megahertz to a few gigahertz~\cite{XueOpt.Express2016,MoilleAPLPhotonics2022a}. Two main effects can be leveraged from thermo-refractive actuation. First, a change in dispersion occurs since the effective refractive index, being frequency-dependent, arises from the difference in confinement with frequency, where the ring core and cladding materials exhibit different thermo-refractive coefficients. Hence, the FSR may change~\cite{MoilleAPLPhotonics2022a}. Second, it alters the detuning of the main pump from the cavity, which may result in a change in the repetition rate, particularly in the presence of asymmetric dispersion or the Raman effect. Regardless, the maximum speed for thermal tuning is typically limited to the ten kilohertz range since the thermal lifetime of the integrated system is slow. This speed is far too limiting in controlling the terahertz comb. On the other hand, electro-optic~\cite{HeNatCommun2023} and piezo-electric effects~\cite{LiuNature2020} can provide fast tuning, with on-chip EO modulators demonstrated with bandwidths exceeding 100~GHz~\cite{WangAPLPhotonics2019}. In this case, the dispersion remains mostly intact, negligibly impacting $\omega_\mathrm{rep}$. Instead, the cavity resonance frequencies are modulated, which in turn influences the detuning of the main pump, thereby causing a variation in the repetition rate. In these devices, repetition rate tuning on the order of a few megahertz can be achieved \cite{HeNat.Commun.2023,LiuNature2020}, which is similar to the tens of megahertz achievable through KIS. Unlike the EO effect, the piezoelectric effect can only efficiently modulate the resonator at fixed mechanical frequencies determined by its size. This is quite limiting as well for terahertz voltage-controlled-oscillator functionality. In addition, even though these three different methods are based on different physical effects, they all rely on the actuation of the cavity within which the DKS lives, by tuning its resonance frequencies. Hence, the overall excursion is limited by the DKS pump frequency detuning existence window, since a large frequency tuning would result in loss of the cavity soliton.

In sharp contrast, the KIS scheme demonstrated here does not rely on cavity actuation, but rather direct actuation of the group velocity of the cavity soliton (without impacting the cavity). This is a key difference since there is no impact on the DKS existence window (since the pump frequency is unchanged), while repetition rate actuation relies only on intracavity nonlinear dynamics instead of increased complexity in the device architecture for on-chip electrodes or incorporation of electrically/mechanically-active materials. The overall speed at which the DKS can be modulated is only limited by the cavity linewidth, which without further optimization has an upper limit of about 1~GHz, which is on par with EO and piezo-based systems for DKS modulation. This bandwidth can be improved through coupling engineering to reduce the photon lifetime, which has been shown to significantly increase the KIS bandwidth~\cite{Moille2024_colorKIS}. 

                                                 
\section{Conclusion} 
In conclusion, we have introduced and experimentally demonstrated a new concept to create an integrated terahertz voltage-controlled-oscillator that transduces a microwave signal carried by an electro-optically modulated reference laser into the the terahertz regime, through dynamic Kerr-induced synchronization of a cavity soliton. We have shown its ability to arbitrarily tune frequency around the DKS repetition rate near 1~THz, yielding a terahertz comb with 13~MHz bandwidth and consisting of about 130 comb teeth that carry the same noise as the DKS repetition rate, which is obtained from the main and reference pump noise through optical frequency division. Our work presents a proof-of-concept approach for a tunable and flexible terahertz oscillator, where different metrics -- while already on par with other techniques for modulating DKS microcombs --  could be readily improved by locking the two pumps for better noise performance or coupling engineering for faster modulation speed. Moreover, our demonstration showcases direct DKS actuation rather than dynamic modification of the cavity in which the DKS lives, bypassing certain critical limitations in DKS-based optical systems for microwave and terahertz signal generation.

\subsection*{Acknowledgments}
The photonic chips were fabricated in the same fashion as those presented in~\cite{MoilleNature2023}. The Scientific colour map batlow~\cite{Crameri2023} and subsequent color set is used in this study to prevent visual distortion of the data and exclusion of readers with colour-vision deficiencies~\cite{CrameriNatCommun2020}. We acknowledge partial funding support from the Space Vehicles Directorate of the Air Force Research Laboratory and the NIST-on-a-chip program of the National Institute of Standards and Technology. We thank David Long and Yichen Shen for insightful feedback.
 
\subsection*{Author Contributions}
U.A.J and M.C. performed the experiment and processed the data. G.M. conceptualized the project and designed the resonators. G.M. and K.S. helped with data processing and the understanding of the phenomenon. K.S. and G.M. led the project and helped with data analysis. All the authors wrote, contributed and discussed the content of this manuscript

\subsection*{Competing Interests}
G.M. and K.S have submitted a provisional patent application based on aspects of the work presented in this paper.

\subsection*{Data availability}
The data that supports the plots within this paper and other findings of this study are available from the corresponding authors upon request.


\clearpage
\onecolumn
\setboolean{singlecolumn}{true}
\renewcommand{\appendixpagename}{
\vphantom{M}%
\vskip-0em%
\raggedright \titlefont Supplementary Informations: \mytitle}
\appendix   
\appendixpage
\renewcommand{\thesection}{S.\arabic{section}}
\renewcommand\thefigure{S.\arabic{figure}}    
\setcounter{figure}{0}

\vspace{1em}
\section{Additional data}
\begin{figure*}[h!]
\centering
  \includegraphics[scale=0.45]{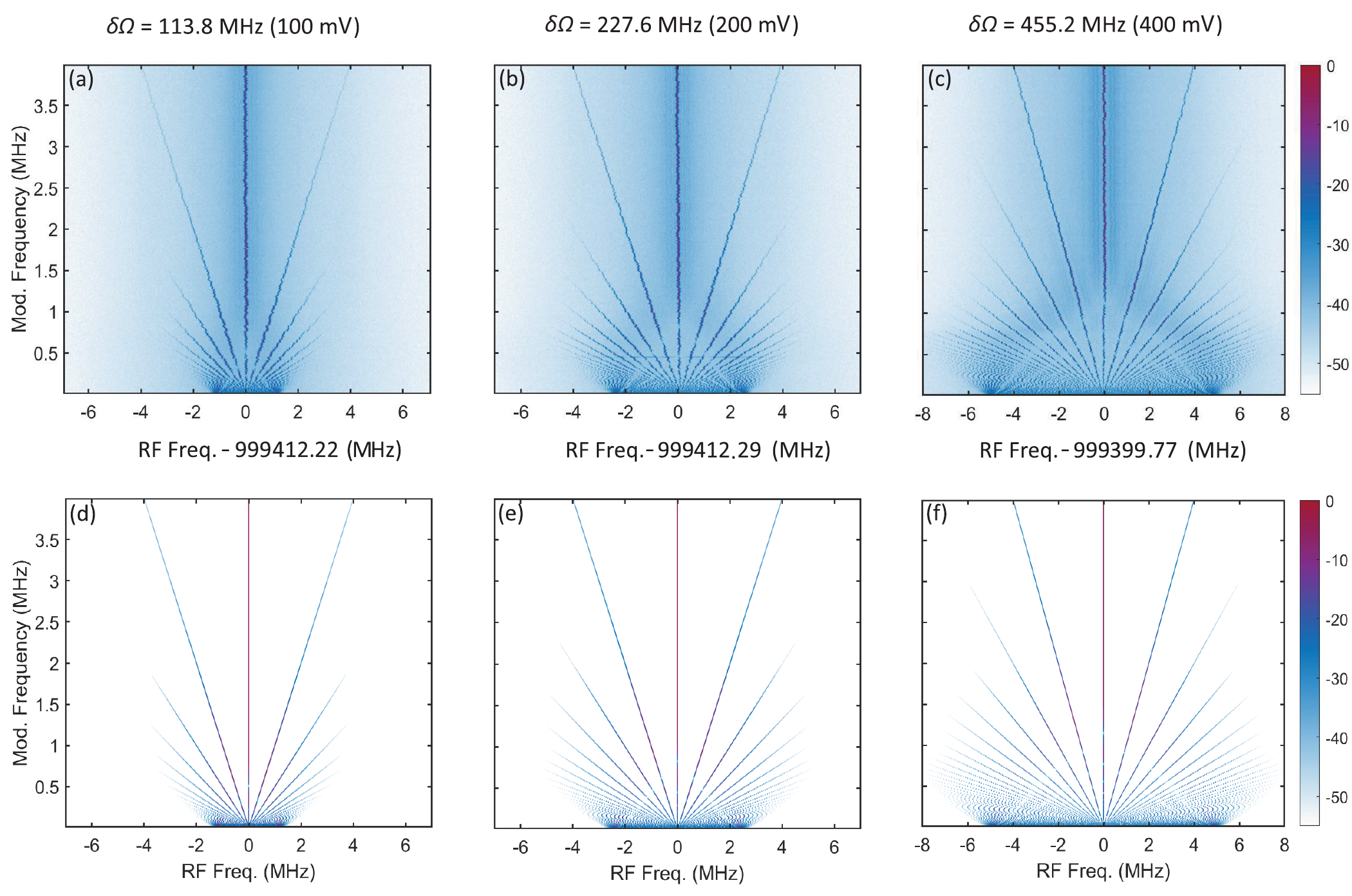}
  \caption{Terahertz frequency comb spectra for three different modulation amplitude values $\delta\Omega$: (a) 113.8 MHz (b) 227.6 MHz, and (c) 455.2 MHz , controlled with peak-to-peak applied voltage to the VCO, with modulation frequency varied from 0.1~MHz to 4~MHz. The offset frequency of the central line is provided with each plot. The corresponding theoretical plots are provided in (d)-(f).}
  \label{figS11}
\end{figure*}

Terahertz comb spectra obtained at different modulation amplitudes are plotted in Fig.~\ref{figS11}, along with theoretical fits. The corresponding center frequencies are also provided.

\begin{figure*}[!h]
\centering
  \includegraphics[scale=1.2]{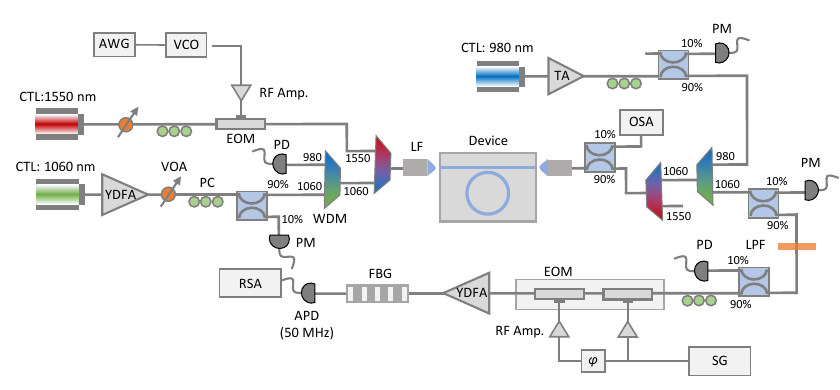}
  \caption{Experimental setup for terahertz comb generation and characterization. Detailed explanation is provided in the accompanying text. CTL: Continuously-tunable laser, YDFA: Ytterbium-doped fiber amplifier, VOA: variable optical attenuator, PC: Polarization controller, AWG: Arbitrary waveform generator, VCO: Voltage-controlled oscillator, PD: Photo-detector, PM: Power meter, WDM: Wavelength-division multiplexer, LF: Lensed fiber, OSA: Optical spectrum analyzer, LPF: Long-pass filter, SG: Signal generator, FBG: Fiber Bragg grating filter, APD: Avalanche photo-diode, RSA: Real-time spectrum analyzer. }
  \label{fig:Exp}
\end{figure*}

\vspace{1em}
\section{Experimental setup and repetition rate measurement}
Figure \ref{fig:Exp} shows the experimental setup. The experiment uses three lasers. One laser pumps the soliton at 1061 nm (CTL) with 140 mW of optical power provided after an Ytterbium-doped fiber amplifier (YDFA) and the second CTL thermally stabilizes the resonator to easily access the soliton detuning regime, by pumping in the opposite direction with orthogonal polarization at 981 nm, with 350 mW of optical power provided after a tapered amplifier (TA). A third laser is used to synchronize a soliton comb tooth at 1566 nm with $\approx~$0.2 mW on-chip power in the first sideband of the EO modulator. These lasers are combined using fiber-optic wavelength division multiplexers (WDMs) and sent into lensed fibers that couple the light into the chip. At the output, 10~\% of light is tapped out and sent to an optical spectrum analyzer (OSA) to capture the comb spectrum and the rest is split using WDMs to reject the lasers. A section of the comb spectrum on the red side of the pump laser is filtered out using a free-space long-pass filter at 1070 nm.

\begin{figure*}[t]
\centering
  \includegraphics[scale=1]{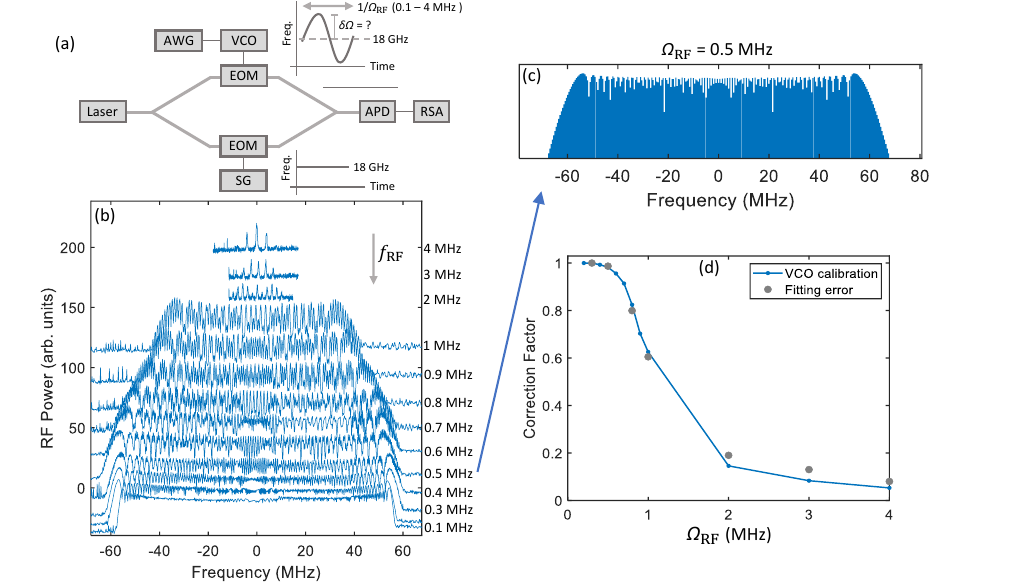}
  \caption{Calibration of the VCO excursion. (a) A frequency shearing Mach-Zehnder interferometer with the VCO driven EO modulator in one arm and a second EO modulator driven by a second signal generator (SG). The modulation signal in the lower arm is a single frequency 18 GHz sine wave while the VCO applies a frequency modulated signal with the same 18 GHz center frequency and an excursion $\delta\Omega$ at a frequency $\Omega_{\mathrm{RF}}$ varied from 0.1 MHz to 4 MHz. (b) Measured beat signal spectra at the output of the interferometer at different values of $\Omega_{\mathrm{RF}}$. (c) A sample theoretical fit of the generated beat note spectrum at $\Omega_{\mathrm{RF}}=$ 0.5 MHz. These fits are used to determine the value of $\delta\Omega$ at each modulation frequency. (d) The theoretical fits of the data in (b) provide a correction factor for $\delta\Omega$ at each modulation frequency, shown in blue.
  }
  \label{figS1}
\end{figure*}

As mentioned in the main article, the 1~THz repetition rate was indirectly measured by down-shifting it to the megahertz domain and using a fast optical detector. These modulators are driven at a frequency of 17.84706 GHz and close to saturation with a power of 3~W (35 dBm) using RF amplifiers. The relative phase between the modulators is adjusted to maximize the bandwidth of the generated EO comb. The modulation frequency is chosen such that the 56th multiple of this number approximately matches the $\approx 1$~THz repetition rate. This guarantees that one comb line from each side overlaps in the middle, with a difference small enough to be picked up with an InGaAs APD with 50~MHz bandwidth. The spectrum in this overlap region is filtered out with a 0.1 nm tunable bandpass fiber bragg grating filter (FBG) and fed to the APD. The output of the APD is sent to a real-time spectrum analyzer (RSA) which gives the down-shifted terahertz frequency comb spectra.

\vspace{1em}
\section{VCO calibration}
The VCO used in the experiment has a response time limited to a bandwidth of 0.5 MHz, after which its excursion starts to drop. This makes the effective modulation amplitude smaller at a given applied voltage at frequencies larger than 0.5 MHz, and is observed in the measured excursion of the DKS repetition rate, which reduces drastically at high frequencies. Therefore, a calibration of the VCO excursion for each frequency is needed. To do this, we set up a frequency shearing Mach-Zehnder interferometer (MZI), with the VCO in one arm and another microwave signal generator (SG) in the other arm, as shown in Fig.~\ref{figS1}(a). The signal generator is driven at a frequency of 18 GHz, the same used in the experiment as the separation between the reference laser central line and its first EO sideband. The VCO signal is swept with the same 18 GHz center frequency and an amplitude corresponding to a sine wave with a peak-to-peak amplitude of 50 mV, applied using the AWG. The measurement then consists of determining the VCO modulation amplitude $\delta\Omega$ corresponding to this voltage. To do so, the frequency of the sine wave $\Omega_{\mathrm{RF}}$ is varied from 0.1 MHz to 4 MHz, and a telecom band CW laser is sent into one input of the interferometer. The light is modulated in the two arms and then recombined at the output. This mixing downshifts the EO modulation imparted by the VCO to DC, where we can measure the frequency excursion with a 150 MHZ bandwidth APD. The output of the APD is fed into the RSA. Figure~\ref{figS1}(b) plots the measured spectra from the beating of the SG and VCO revealing the exact excursion for varying modulation frequency. The MZI output produces an EO comb corresponding to the Fourier transform of a frequency modulated sine wave at 18 GHz down-shifted to DC. Its bandwidth can be measured and fitted to obtain the modulation amplitude that produces an identical spectrum. This will give us the modulation amplitude $\delta\Omega$ corresponding to the 50 mV signal we applied to the VCO. Immediately, we can see that the bandwidth of the EO combs remains the same up till 0.5 MHz after which it slowly starts to decline. After 1 MHz the decline is drastic. Using the same Bessel function expansion of a frequency modulated signal as we did in the main article, we fit each of these spectra. Below 0.5 MHz, the VCO excursion maintains its value, which we fit to obtain a modulation amplitude $\delta\Omega=$ 59.6 MHz. Using this, we obtained the $\delta\Omega=$ 341.4 MHz for the terahertz comb plot in Fig.~4(a) in the main article, which was taken at a 300 mV peak-to-peak sine wave. For the plots in Fig.~\ref{figS11}, the modulation amplitudes are 113.8 MHz for 100 mV, 227.6 MHz for 200 mV, and 455.2 MHz for 400 mV peak-to-peak sine waves. For the smallest excursion at 100 mV in Fig.~\ref{figS11}(a), the fitted modulation amplitude of the repetition rate is evaluated as $\delta\Omega_{\mathrm{rep}}=$ 1.2504 MHz, giving us a ratio of 1.2504/113.8=1/91.011. This is in excellent agreement with the OFD factor $|\Delta\mu|$=1/91. For the larger modulation amplitudes (200 mV, 300 mV and 400 mV) in Fig.~\ref{figS11}(b),(c), the fitted modulation amplitudes remain within an error of 3.3\% from the OFD factor. 
\par The fitted modulation amplitudes for each frequency give us a frequency dependent correction factor to apply in our theoretical model, which we have used to obtain the theoretical plots in Fig.~4(c) in the main article and in Fig.~\ref{figS11} here. This correction factor vs frequency is plotted in Fig.~\ref{figS1}(d) and labelled as VCO calibration. To verify that the frequency response is solely responsible for the reduced excursion of the repetition rate, we have calculated a fitting error for the measured therahertz comb spectra for Fig.~\ref{figS11}(a) without any correction applied at several frequencies. These are also shown in Fig.~\ref{figS1}(d) along with the VCO calibration. The two data agree well, indicating that this is indeed the source of the reduced frequency response.

\end{document}